\documentclass[11pt]{article} 

\usepackage[margin=1in]{geometry}
\usepackage{amsmath,amssymb}
\usepackage{bm}
\usepackage{graphicx}
\usepackage[utf8]{inputenc}
\usepackage[T1]{fontenc}
\usepackage[numbers,super,sort&compress]{natbib}
\usepackage[colorlinks=true,linkcolor=blue,citecolor=blue,urlcolor=blue]{hyperref}
\usepackage{setspace}
\usepackage{booktabs,array}
\usepackage{placeins}

\graphicspath{{figs/}{.}}

\newcommand{\gdbc}{GdB$_3$C$_3$}
\newcommand{\Ueff}{U_{\mathrm{eff}}}
\newcommand{\Ef}{E_{\mathrm{F}}}
\newcommand{\kv}{\mathbf{k}}

\begin{document}
\title{\textbf{Cage-enforced $i$-wave altermagnetism in boron--carbon clathrates}}
\author{Meiyan Wang$^{1}$, Li Zhu$^{1,\ast}$\\[6pt]
\normalsize $^{1}$Department of Physics, Rutgers University, Newark, New Jersey 07102, USA\\
\normalsize $^{\ast}$Corresponding author: li.zhu@rutgers.edu}
\date{}
\maketitle

\begin{abstract}
\noindent
The spin splitting of an altermagnet arises from a rotation that relates its two magnetic
sublattices. In known cases it is fixed by the crystal structure. Here we show that guest--host
frameworks separate the two roles. In the boron--carbon clathrates $M$B$_3$C$_3$ the ordered
cage removes the translation between the two guest sites and enforces cubic $i$-wave
altermagnetism, with six symmetry-enforced nodal planes, for any magnetic guest in the
compensated state; strain retunes the class to $d$- or $g$-wave. We demonstrate the platform
with \gdbc{}: its compensated antiferromagnetic ground state is a narrow-gap semiconductor
with spin splitting of 0.19~eV at the conduction-band edge, and its weakly pinned N\'eel vector
switches the helicity-resolved optical response on and off according to the magnetic point
group. Thermodynamic stability at high pressure and dynamical stability at ambient pressure
indicate that it can be made by the route established for this family.
\end{abstract}

\section*{Introduction}

Altermagnets of a given class have so far been found one compound at a time, because the rotation that relates their two magnetic sublattices, and with it the class, is a property of each crystal structure. Here we show that a nonmagnetic cage can supply that rotation for any guest. Altermagnetism is a collinear magnetic order that combines compensated moments with momentum-dependent spin splitting~\cite{Smejkal2022PRX1,Smejkal2022PRX2,Jungwirth2026signatures}.
The splitting exists without spin--orbit coupling
(SOC)~\cite{Hayami2019momentum,Yuan2020giant,Smejkal2020crystal}
because the two oppositely polarized sublattices are connected by a
crystallographic rotation rather than by a translation or an
inversion, as formalized in the spin-space-group
description~\cite{Liu2022spin,Xiao2024spin}.
That rotation also fixes the angular symmetry of the splitting,
$d$-, $g$- or $i$-wave, and the planes on which it must
vanish~\cite{Smejkal2022PRX1,Ezawa2025PRB}.
Spin splitting has been resolved in MnTe and
CrSb~\cite{Krempasky2024,Lee2024,Reimers2024,yang2025three},
altermagnetic domains have been imaged and
manipulated~\cite{Amin2024nanoscale,Zhou2025manipulation}, and the
early metallic prototype RuO$_2$ remains
debated~\cite{Feng2022anomalous,Hiraishi2024nonmagnetic,Liu2024absence}.
The established cases are of $d$- and $g$-wave type. The highest
class, $i$-wave, has the cubic form
\begin{equation}
F_6(\mathbf{k}) =
(k_x^2-k_y^2)(k_y^2-k_z^2)(k_z^2-k_x^2),
\label{eq:f6intro}
\end{equation}
which vanishes on six symmetry-related nodal planes. It has been
proposed in rhombohedral compounds from a machine-learning
search~\cite{Gao2025}, in a two-dimensional
monolayer~\cite{mohanta2026wave} and in a minimal model for the
$Pm\bar{3}n$ space group~\cite{Roig2024}, and the cubic rare-earth
compound Ce$_4$Sb$_3$ has been predicted to be a $d$-wave
altermagnet~\cite{He2025ce4sb3}. To our knowledge no bulk material with cubic $i$-wave altermagnetism has been reported, although the higher-order classes have distinct transport signatures~\cite{Ezawa2025PRB}.

Guest--host frameworks separate the two ingredients.
The magnetic ions occupy the guest sites, while the nonmagnetic host
determines which operations relate neighboring guests and therefore
whether a compensated order is altermagnetic, and of which
class~\cite{JaeschkeUbiergo2024supercell}. Magnetically ordered guest--host compounds are known, for example the ferromagnetic $4f^{7}$ clathrate Eu$_8$Ga$_{16}$Ge$_{30}$~\cite{Paschen2001eu8ga16ge30}, and the
same separation of roles underlies the predicted ferroelectricity of
the clathrate ScB$_3$C$_3$~\cite{Zhu2020ferroelectric}; altermagnetism,
however, has not been identified in any guest--host compound. The
boron--carbon clathrates
$M$B$_3$C$_3$~\cite{Zhu2020SciAdv,Strobel2021lanthanum,Wang2021high}
realize this separation in a simple form. They adopt the A15-type cubic
framework of space group $Pm\bar{3}n$, in which a covalently bonded,
ordered B--C network encloses the guests.
SrB$_3$C$_3$~\cite{Zhu2020SciAdv,Zhu2023superconductivity}, the
trivalent-guest member LaB$_3$C$_3$~\cite{Strobel2021lanthanum} and a
type-I boron--carbon clathrate in the same space
group~\cite{Strobel2025typeI} have been synthesized at high pressure
and recovered to ambient conditions. The rare-earth members have
been predicted to host heavy-fermion and Kondo-insulating
states~\cite{Rao2023heavy}. The two guest sites form a
CsCl-like arrangement that, on its own, would be connected by the
body-centring translation. The ordered B--C cage removes this
translation while keeping the rotations that exchange the guests,
and, as shown below, it does so because of the guest site symmetry alone. Any magnetic guest on this site in the two-sublattice N\'eel
state is therefore a cubic $i$-wave altermagnet, strain retunes the
class to $d$- or $g$-wave, and whether a given member is altermagnetic depends only on its magnetic ground state.

Within the family, GdB$_3$C$_3$ is the first member to examine. Gd$^{3+}$ orders antiferromagnetically in the boron cage of
GdB$_6$~\cite{Geballe1968magnetic,Kunii1985gdb6}, and its half-filled
$4f^{7}$ shell ($S=7/2$, $L=0$) suppresses the leading contribution to
magnetocrystalline anisotropy~\cite{ColarietiTosti2003origin}, so that
the N\'eel vector is only weakly coupled to the lattice. This is relevant to the optical response of altermagnets, where the magnetic point
group of a given N\'eel-vector orientation decides which antisymmetric
responses are
allowed~\cite{Smejkal2020crystal,Zhou2021crystal,Hariki2024xray,Kunes2025neel};
a rotatable N\'eel vector therefore allows these selection rules to be switched within one compound. Rare-earth altermagnets have been proposed, among them TbPt$_6$Al$_3$~\cite{Oishi_2025} and the Gd compounds GdAlSi and GdAlGe~\cite{Nag2024gdalsi,Parfenov2025competing}. Resonant
x-ray scattering has since shown the ground state of GdAlSi to be a
spin cycloid rather than a collinear
altermagnet~\cite{Nakano2026cycloid}. The combination of localized
$4f$ moments with a framework that dictates the altermagnetic class
has not been explored.

Here we establish the boron--carbon clathrates as a platform for cubic
$i$-wave altermagnetism and demonstrate it for \gdbc{} by magnetic
symmetry analysis and first-principles calculations. We show that the
framework enforces the altermagnetic class for every guest. For
\gdbc{}, the compensated state is a narrow-gap semiconductor with spin
splitting reaching 0.19~eV near the conduction-band edge and six
symmetry-enforced nodal planes, its magnetocrystalline anisotropy is of order $\mu$eV, and helicity-resolved interband transitions switch on and off with the N\'eel-vector orientation as the magnetic point group requires. Thermodynamic stability at high pressure and dynamical stability at ambient conditions, in a family already synthesized by the same route, indicate that the compound can be made.

\section*{Results}

\subsection*{Crystal structure and altermagnetic symmetry}

\gdbc{} crystallizes in the cubic A15-type structure with
space group $Pm\bar{3}n$ (No.~223). The unit cell contains
two Gd, six B, and six C atoms, i.e.\ two formula
units of \gdbc{}. Full structural relaxation
at ambient pressure yields a lattice constant of
$a=4.8175$~\AA. The Gd atoms occupy the positions $(0,0,0)$ and
$(\tfrac12,\tfrac12,\tfrac12)$, each surrounded by a
B$_{12}$C$_{12}$ cage of the ordered boron--carbon framework,
with B and C atoms occupying the $6c$ and $6d$ Wyckoff sites,
respectively (Supplementary Fig.~1).

\begin{figure}[tbp]
\centering
\includegraphics[width=0.88\textwidth]{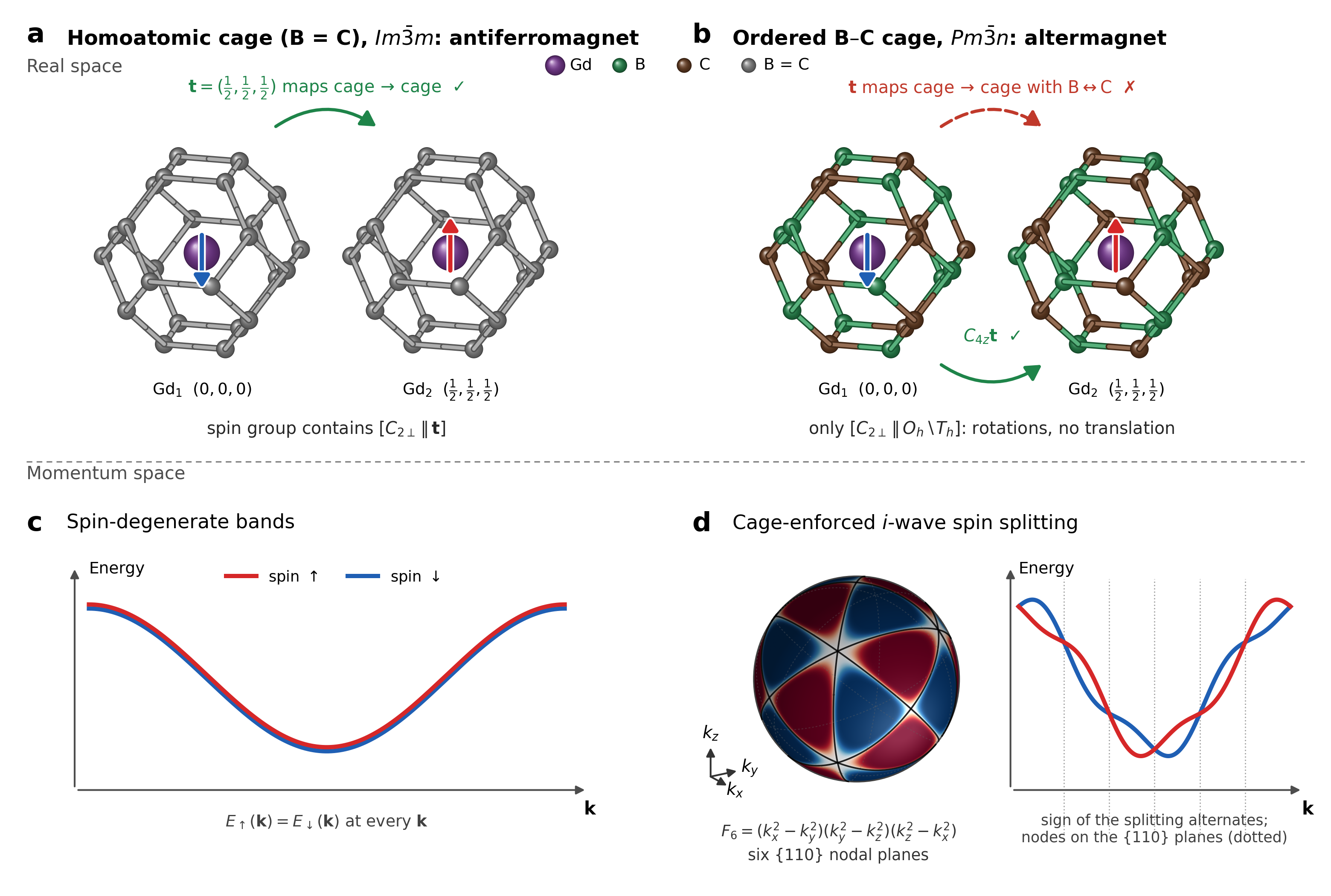}
\caption{\label{fig:structure}
\textbf{Cage-enforced $i$-wave altermagnetism.}
\textbf{a}~Reference structure with a homoatomic cage (B = C, $Im\bar{3}m$; hypothetical,
for symmetry comparison only). The two guests at $(0,0,0)$ and
$(\tfrac12,\tfrac12,\tfrac12)$ sit in identical cages related by the body-centring
translation $\mathbf{t}=(\tfrac12,\tfrac12,\tfrac12)$, so the spin group contains
$[C_{2\perp}\,\|\,\mathbf{t}]$ and the N\'eel state is a conventional antiferromagnet.
\textbf{b}~The ordered B--C cage of $Pm\bar{3}n$ \gdbc{}. The translation $\mathbf{t}$ maps
the cage of Gd$_1$ onto that of Gd$_2$ with B and C exchanged and is therefore lost; the
two cages, and hence the two antiparallel moments (red and blue arrows), are related only by
the 24 rotations of $O_h\setminus T_h$, such as $C_{4z}\mathbf{t}$. Purple, Gd; green, B;
brown, C; grey, the homoatomic cage of \textbf{a}. Cage geometry from the relaxed structure.
\textbf{c}~Spin-degenerate bands of the conventional antiferromagnet (schematic).
\textbf{d}~Sign of the $i$-wave form factor
$F_6=(k_x^2-k_y^2)(k_y^2-k_z^2)(k_z^2-k_x^2)$ on a sphere in momentum space (red positive,
blue negative), with the six $\{110\}$ nodal planes, and a schematic of bands whose spin
splitting changes sign at each nodal plane.}
\end{figure}

The two Gd sites form a CsCl-like arrangement, but the translation $\bm{t}=(\tfrac12,\tfrac12,\tfrac12)$ that would connect them in the isolated Gd sublattice exchanges the B and C sites and is not a symmetry of the crystal (Fig.~\ref{fig:structure}a,b). The two Gd sublattices are instead related by 24 proper and improper rotations combined with fractional translations; none is a pure translation, and inversion, although present, maps each sublattice onto itself. When the two sublattices carry equal and opposite moments, no symmetry enforces spin degeneracy throughout the Brillouin zone, and the compensated state is an altermagnet with momentum-dependent spin splitting even without spin--orbit coupling. This classification follows from symmetry alone and is independent of the Hubbard parameter and the exchange--correlation functional.

The result is a property of the framework rather than of Gd. The
guest site $2a$ of $Pm\bar{3}n$ has site symmetry $m\bar{3}$ ($T_h$,
order 24), a subgroup of index two of the point group $m\bar{3}m$
($O_h$, order 48). The 24 operations that exchange the two guests are
exactly the coset $O_h \setminus T_h$: the six fourfold rotations, the
six twofold rotations about $\langle110\rangle$, the six $S_4$
rotoreflections and the six diagonal mirrors, each combined with the
translation $(\tfrac12,\tfrac12,\tfrac12)$. Since the identity and
the inversion both belong to $T_h$, neither can exchange the
sublattices, whatever the guest element, the lattice parameter or the
decoration of the cage. Every magnetic guest on the $2a$ site of a
$Pm\bar{3}n$ clathrate in the two-sublattice N\'eel state is therefore
a symmetry-eligible altermagnet of the same class. This includes the
small-cage guests of the type-I boron--carbon
clathrate~\cite{Strobel2025typeI}. Whether a given member is an altermagnet therefore depends only on its magnetic ground state. Lowering the symmetry does not remove the altermagnetism but
changes its class. A tetragonal strain along a cubic axis
($P4_2/mmc$) leaves a $d$-wave altermagnet and a rhombohedral strain
along $[111]$ ($R\bar{3}c$) a $g$-wave one (Supplementary Note~4),
so the wave class of a single compound is, in principle,
strain-switchable. We next determine the magnetic ground state and the electronic structure of the Gd member.

\subsection*{Magnetic ground state and electronic structure}

Across the lanthanide series the ferromagnetic and compensated states are nearly degenerate, and Gd is the member that tips to the compensated side. We surveyed the series $M$B$_3$C$_3$ by comparing ferromagnetic (FM) and antiferromagnetic (AFM) configurations within collinear $r^{2}$SCAN$+U$ calculations.
At $\Ueff = 4.2$, 5.5 and 7.0~eV every member that converges
reliably favors ferromagnetic order except Gd, which crosses to the
compensated state between 5.5 and 7.0~eV; Ce, Pr and Er remain within
1~meV of degeneracy at every $\Ueff$ (Supplementary Table~1). The pattern mirrors the hexaborides, in which Gd$^{3+}$ in GdB$_6$ orders
antiferromagnetically below 15--16~K whereas Eu$^{2+}$ in EuB$_6$
is a ferromagnet~\cite{Geballe1968magnetic,Kunii1985gdb6,Amara2005exchange,Sullow1998structure}.
In a gapped host such as \gdbc{} the coupling is expected to be
of superexchange type through the covalent framework, rather than
the carrier-mediated RKKY interaction of the metallic
hexaborides~\cite{Anderson1963theory,Ruderman1954indirect,Kasuya1956theory,Yosida1957magnetic}.

For \gdbc{} the outcome depends on the on-site Coulomb interaction of the Gd $4f$ states. $r^{2}$SCAN$+U$ gives a crossover from FM to AFM order near $\Ueff \approx 6.1$~eV, and at $\Ueff = 7.0$~eV the compensated AFM state lies below the FM configuration by $\Delta E = E_{\mathrm{FM}}-E_{\mathrm{AFM}} = 1.99$~meV per magnetic unit cell (two Gd), rising to 14.1~meV at $\Ueff = 10$~eV (Supplementary Table~2). The compensated state is thus the ground state, but only by about 1~meV per Gd, a near-degeneracy characteristic of Gd$^{3+}$ compounds with weak inter-site coupling~\cite{Parfenov2025competing,Mitra2008magnetic}.

The value of $\Ueff$ is anchored in two independent ways. First, a self-consistent constrained random-phase
approximation
(cRPA)~\cite{Aryasetiawan2006calculations,Nilsson2013abinitio}
for the Gd $4f$ shell of \gdbc{} itself converges to $U = 7.51$~eV
and $J = 0.23$~eV, i.e.\ $\Ueff = U - J = 7.28$~eV (Methods).
This agrees with the $U = 8.2$~eV obtained ab initio for Gd$^{3+}$
in the wide-gap insulator GdScO$_3$~\cite{Mizzi2018} and with the
parameters established for Gd in the rare-earth
nitrides~\cite{Larson2007electronic}. Second, the same functional
and $\Ueff$ reproduce the antiferromagnetic ground state of
GdB$_6$~\cite{Geballe1968magnetic,Kunii1985gdb6,Xu2019gdb6}
by 13--14~meV per Gd for $\Ueff = 5$--7~eV
(Supplementary Table~3). The empirical $\Ueff = 7.0$~eV adopted for
all electronic-structure calculations below is therefore within
0.3~eV of the first-principles value.

The altermagnetic classification, the spin splitting and the optical selection rules established below are properties of the compensated configuration.

\begin{figure}[tbp]
\centering
\includegraphics[width=\textwidth]{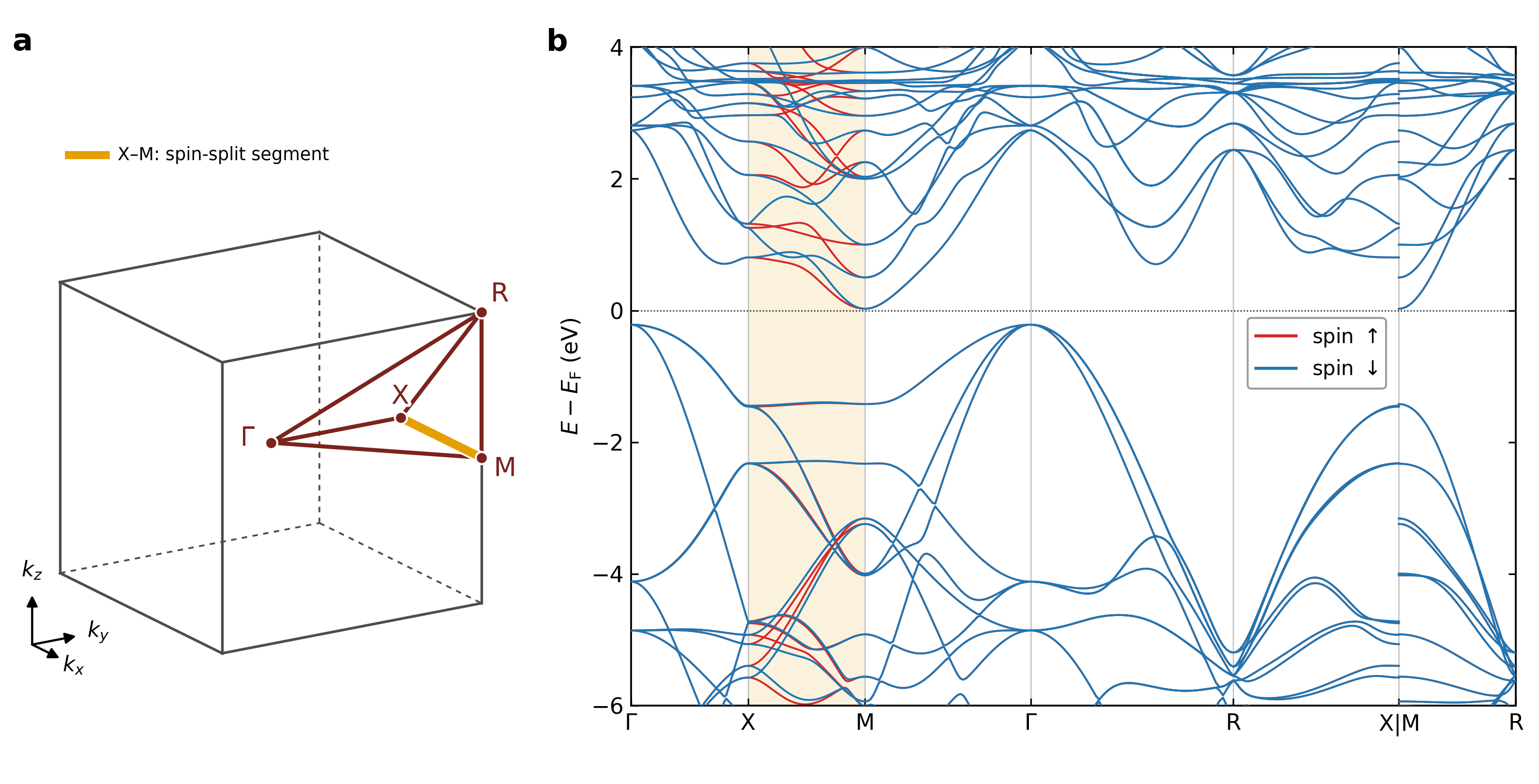}
\caption{\label{fig:bands}
\textbf{Spin-resolved band structure of the compensated antiferromagnetic state.}
\textbf{a}~Simple-cubic Brillouin zone with the band path, $\mathrm{X}=(0,\tfrac12,0)$,
$\mathrm{M}=(\tfrac12,\tfrac12,0)$, $\mathrm{R}=(\tfrac12,\tfrac12,\tfrac12)$. X--M, the
spin-split segment of the path, is shown in orange and shaded in \textbf{b}.
\textbf{b}~Electronic band structure of \gdbc{} in the compensated
antiferromagnetic configuration, calculated
using $r^{2}$SCAN$+U$ with
$\Ueff = 7.0$~eV in the absence of
spin--orbit coupling.
Red and blue curves represent the two
spin channels; the zero of energy is the Fermi level of the
calculation, which lies at the conduction-band minimum at M, with the
valence-band maximum at $\Gamma$. The bands exhibit
symmetry-enforced degeneracies along
selected momentum directions and
spin splitting at generic crystal
momenta. The flat unoccupied Gd $4f$ states lie about 3.4~eV above the Fermi level. The path
is discontinuous at X$|$M.
}
\end{figure}

The electronic structure of the compensated AFM
configuration is shown in Fig.~\ref{fig:bands}.
The $r^{2}$SCAN$+U$ calculations predict a
narrow-gap semiconducting state, with an indirect
gap of 0.24~eV between the valence-band maximum at
$\Gamma$ and the conduction-band minimum at M (Fig.~\ref{fig:bands});
the three-dimensional $15\times15\times15$ grid used below finds no
lower band edge. The minimum direct gap is larger, 1.48~eV, from both planar and three-dimensional $k$-point sampling. This distinction is relevant
to the optical response discussed below, since
vertical interband optical transitions are governed
by the direct excitation energies rather than
the indirect fundamental gap.

The electronic states near the band edges are
predominantly derived from B/C-$p$ and Gd-$d$
orbitals, whereas the flat unoccupied Gd $4f$ manifold lies about 3.4~eV above the Fermi level. The spin splitting near the band edges therefore involves dispersive states rather than the localized $4f$ bands. As shown in
Fig.~\ref{fig:bands}, the two spin channels
remain degenerate along symmetry-constrained
directions but separate at general crystal
momenta, consistent with the altermagnetic
symmetry established above. Its momentum dependence is examined next.

\subsection*{Symmetry-enforced $i$-wave spin splitting}

Symmetry requires the spin splitting of the compensated state to vanish on six planes, and the calculation finds that it does, to one part in $10^{4}$. Without spin--orbit coupling the spin-up and spin-down states form two independent channels, so the signed spin splitting can be defined as
\begin{equation}
    \Delta_n(\mathbf{k})
    =
    E_{n\uparrow}(\mathbf{k})
    -
    E_{n\downarrow}(\mathbf{k}),
    \label{eq:delta}
\end{equation}
where $n$ denotes the band index. A nonzero
$\Delta_n(\mathbf{k})$ therefore reflects the nonrelativistic
spin splitting associated with the compensated magnetic order,
rather than a relativistic SOC effect. The symmetry of this
splitting is determined by the crystallographic operations
that exchange the two oppositely polarized Gd sublattices.

The spin-group symmetry of \gdbc{} constrains the momentum dependence of the nonrelativistic spin splitting~\cite{Liu2022spin,Xiao2024spin}.
Among the 48 spatial operations of $Pm\bar{3}n$, 24 exchange the
two oppositely polarized Gd sublattices. Six of these operations have $\{110\}$ reflection components and
correspond to glide reflections in the nonsymmorphic space group.
Their accompanying fractional translations exchange the two Gd
sublattices, while the reflection components determine the
transformation of crystal momentum. The spin splitting therefore changes sign under these operations and must vanish
on the six symmetry-enforced nodal planes,
\begin{equation}
\begin{aligned}
    k_x &= \pm k_y,\\
    k_y &= \pm k_z,\\
    k_z &= \pm k_x.
\end{aligned}
\label{eq:nodal_planes}
\end{equation}
By contrast, the $\{100\}$ mirror operations preserve the
individual Gd sublattices and therefore do not enforce spin
degeneracy on the coordinate planes $k_x=0$, $k_y=0$, or
$k_z=0$. This distinction fixes the nodal structure of the cubic $i$-wave splitting.

To identify the lowest-order symmetry-allowed form,
we construct homogeneous polynomials in
$(k_x,k_y,k_z)$ that remain invariant under
sublattice-preserving operations and change sign
under sublattice-exchanging operations. No nonzero
polynomial satisfying these conditions exists through
fifth order. At sixth order, the symmetry constraints
admit a unique polynomial, up to an overall
multiplicative constant,
\begin{equation}
    F_6(\mathbf{k})
    =
    (k_x^2-k_y^2)
    (k_y^2-k_z^2)
    (k_z^2-k_x^2).
    \label{eq:iwave}
\end{equation}
The sixth-degree momentum dependence and the six
symmetry-enforced nodal planes establish the cubic
$i$-wave character of the compensated magnetic
configuration. Equation~\eqref{eq:iwave} describes
the leading symmetry-allowed form near the Brillouin-zone
center; the full DFT spin splitting may additionally
contain higher-order contributions with the same
symmetry.

We verify this classification directly using
three-dimensional first-principles calculations
on a $15\times15\times15$ grid of zero-weight
$k$ points covering the Brillouin zone. The
calculated splitting vanishes to numerical precision
on the symmetry-enforced nodal planes, while
remaining finite at generic momenta. For the
conduction-band manifold represented by band 40,
the maximum absolute splitting on the planes
$k_x=k_z$, $k_x=-k_z$, and $k_y=k_z$
is 0.006, 0.009, and 0.004~meV, respectively,
compared with a maximum of 189~meV
over the full three-dimensional grid.
The corresponding valence-band manifold (band 39)
exhibits a maximum splitting of 14.30~meV,
while its splitting on the same nodal planes
remains below 0.01~meV.
This contrast of four orders of magnitude confirms the predicted nodal structure.

\begin{figure}[tbp]
\centering
\includegraphics[width=0.8\textwidth]{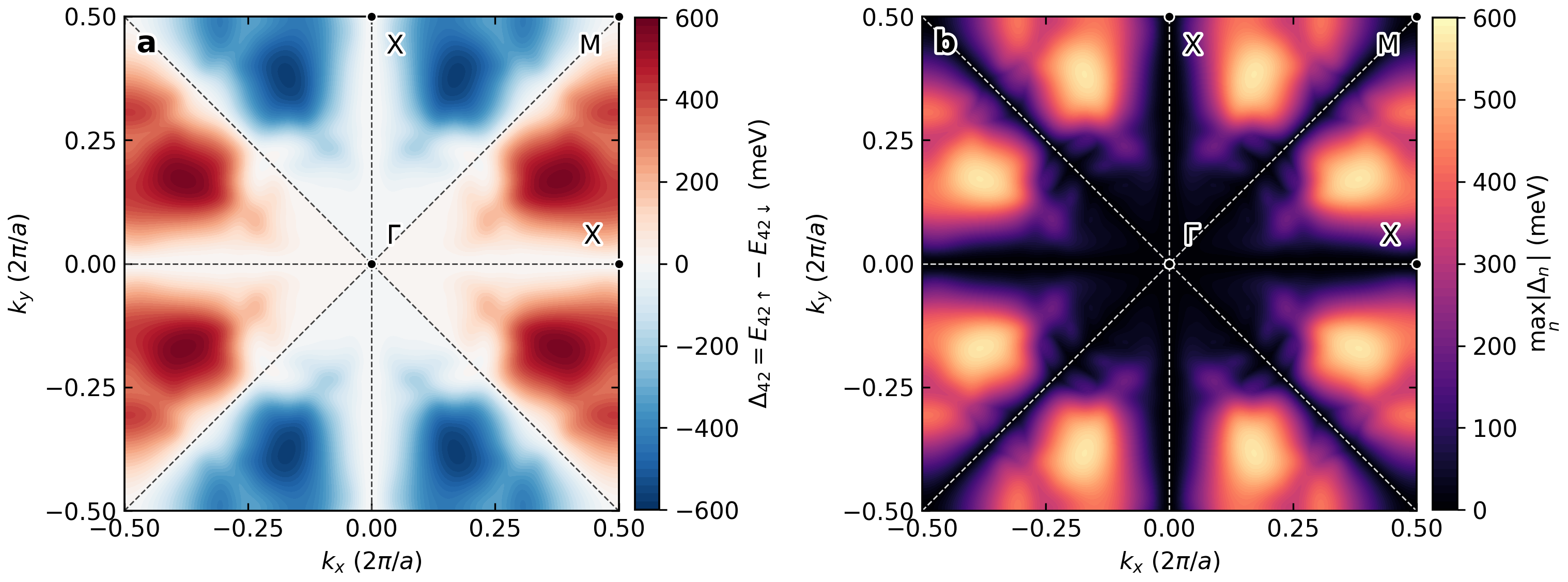}
\caption{\label{fig:map2d}
\textbf{Momentum-dependent spin splitting in the $k_z=0$ plane.}
Spin splitting of \gdbc{} in the $k_z=0$ plane, calculated without
spin--orbit coupling using a $31\times31$
zero-weight $k$-point grid at
$\Ueff=7.0$~eV.
\textbf{a}~Signed splitting $\Delta_{42}(\mathbf{k})$
of band 42 ($\approx 1.95$~eV above $E_{\mathrm F}$),
the band with the largest splitting near the
gap. \textbf{b}~$\max_n|\Delta_n(\mathbf{k})|$ over
bands 36--43 (the four highest valence and
four lowest conduction bands).
Both panels are bicubic interpolations of the
$31\times31$ grid; dashed lines mark the four
nodal lines, and the small cross-shaped features on the diagonals
in \textbf{b} are interpolation artifacts where the band carrying
the maximum changes.
The splitting exhibits
symmetry-enforced nodes along
$k_x=0$, $k_y=0$, and $k_x=\pm k_y$.
These four nodal lines are the intersections
of the six three-dimensional $\{110\}$
nodal planes with the $k_z=0$ plane.
The magnitude of the splitting remains
finite away from the nodal directions.
}
\end{figure}

The splitting in the $k_z=0$ plane shows the same symmetry. The plane contains four nodal lines (Fig.~\ref{fig:map2d}),
$k_x=0$, $k_y=0$, and $k_x=\pm k_y$.
Restricting Eq.~\eqref{eq:iwave} to this plane
gives
\begin{equation}
    F_6(k_x,k_y,0)
    =
    -k_x^2k_y^2(k_x^2-k_y^2).
    \label{eq:iwave_2d}
\end{equation}
Six planes give four lines because the pairs
$k_x=\pm k_z$ and $k_y=\pm k_z$ intersect
the $k_z=0$ plane along $k_x=0$ and $k_y=0$,
respectively.

The nodal pattern in this plane
does not correspond to a conventional planar
$g$-wave form. The factors $k_x^2$ and $k_y^2$
in Eq.~\eqref{eq:iwave_2d} produce zeros
without requiring a sign reversal across
the coordinate axes, whereas the splitting
changes sign across the diagonal nodal lines
$k_x=\pm k_y$.
The three-dimensional symmetry is therefore needed to distinguish cubic $i$-wave splitting from lower-order planar harmonics; the same holds for other three-dimensional altermagnets whose planar cuts resemble a different harmonic~\cite{Urru2025bifeo3}. Consistent with this distinction,
the calculated splitting is orthogonal to
the conventional planar $g$-wave basis
$k_xk_y(k_x^2-k_y^2)$ under the
reported symmetric sampling.

The magnitude of the splitting depends on the band (Table~\ref{tab:delta}). On the three-dimensional grid the valence band 39 reaches 14~meV and the conduction band 40 reaches 189~meV. Higher conduction bands split more, up to 0.57~eV for band 42 in the sampled region.

\begin{table}[tb]
\centering
\caption{\label{tab:delta}
\textbf{Maximum spin splitting of the bands bounding the gap.}
Maximum absolute spin splitting
$\max_{\mathbf{k}}|\Delta_n(\mathbf{k})|$
on the $k_z=0$ plane for selected bands, from the
$31\times31$ in-plane grid (finer in-plane sampling than
the $15\times15\times15$ three-dimensional grid)
near the insulating gap.
The average band energies are measured
relative to the Fermi level.}
\vspace{4pt}
\begin{tabular}{lcc}
\toprule
Band &
$\langle E-E_{\mathrm F}\rangle$ (eV) &
$\max|\Delta_n|$ (meV) \\
\midrule
39 (valence)    & $-0.96$ & 14  \\
40 (conduction) & $+0.97$ & 196 \\
41              & $+1.58$ & 389 \\
42              & $+1.95$ & 567 \\
\bottomrule
\end{tabular}
\end{table}

The largest splitting on the three-dimensional grid within the sampled window, 595.4~meV in band 54 at $\mathbf{k}=(0.500,0.000,-0.143)$, lies on the $k_y=0$ plane, which is not a nodal plane in three dimensions even though its intersection with $k_z=0$, the line $k_y=k_z=0$, is.

The spin-group analysis, the sixth-order invariant and the three-dimensional DFT results together establish the $i$-wave character of the compensated configuration. The much larger splitting of the conduction bands than of the valence bands is a strong electron--hole asymmetry.

\subsection*{Spin--orbit coupling and magnetic anisotropy}

The ground state of the half-filled $4f^{7}$ shell of Gd$^{3+}$ has $S=7/2$ and $L=0$, so the first-order orbital contribution to the magnetic anisotropy vanishes. The magnetocrystalline anisotropy is therefore expected to be weak, with finite contributions only from higher-order SOC effects and from hybridization with other states.

We performed noncollinear
$r^{2}$SCAN$+U$ calculations including SOC for
different N\'eel-vector orientations. The calculated
total-energy differences among the [001], [100],
and [110] configurations are within $17~\mu$eV per magnetic unit cell.
These small energy differences indicate weak
magnetocrystalline anisotropy, consistent with
the predominantly spin-derived magnetic moment
of Gd$^{3+}$~\cite{ColarietiTosti2003origin}. Because the [001]
and [100] orientations are equivalent under the cubic symmetry of
the compensated state, their calculated difference is itself a
measure of the numerical resolution; the energy differences are
comparable to this resolution, and a definitive easy-axis
assignment is not made here.

SOC also changes the dispersion little: the bands with and without SOC are nearly the same near the gap (Supplementary Fig.~2). SOC does, however, set the magnetic point group of each N\'eel-vector orientation, and with it the antisymmetric optical response. A weak anisotropy together with an orientation-dependent magnetic symmetry is the condition under which the N\'eel vector can switch helicity-dependent interband transitions, which we examine next.

\subsection*{N\'eel-vector-dependent optical helicity asymmetry}

Because the anisotropy is weak, the N\'eel-vector orientation is a free parameter of the compensated state. We ask whether reorienting it changes the optical response in the absence of net magnetization, and compute helicity-dependent interband transitions for several orientations.

For circularly polarized light propagating along
the $z$ direction, the transition strengths for
opposite helicities are evaluated from
\begin{equation}
I_{\pm}(\omega)
\propto
\sum_{\mathbf{k},v,c}
\left|
\left\langle
\psi_{c\mathbf{k}}
\left|
\hat{p}_x \pm i\hat{p}_y
\right|
\psi_{v\mathbf{k}}
\right\rangle
\right|^2
G(\hbar\omega-E_{c\mathbf{k}}+E_{v\mathbf{k}}),
\label{eq:optical_helicity}
\end{equation}
where $v$ and $c$ denote occupied and unoccupied
states, respectively, and $G$ is the broadening function. The helicity
asymmetry is characterized by the difference
$I_{+}(\omega)-I_{-}(\omega)$, together with the
total transition strength
$I_{+}(\omega)+I_{-}(\omega)$.

The transition matrix elements are evaluated
using the pseudo-wavefunction momentum operator
within the plane-wave representation. The spectra therefore give relative helicity-dependent transition strengths rather than normalized absorption coefficients, and the implementation omits the projector-augmented-wave augmentation contributions to the optical matrix elements~\cite{Gajdos2006linear}.

The magnetic symmetry of the compensated configuration constrains the antisymmetric optical response for each N\'eel-vector orientation~\cite{Seemann2015symmetry,Smejkal2020crystal,Zhou2021crystal}.
The helicity asymmetry $I_{+}-I_{-}$ for light propagating along
$\hat{\mathbf z}$ is governed by the antisymmetric part of the
dielectric tensor, $\varepsilon_{xy}-\varepsilon_{yx}$, which
transforms as the $z$ component of a time-reversal-odd axial
vector $\mathbf h$. In the collinear compensated state the
sublattice-exchanging operations of $Pm\bar{3}n$ are the elements
of the point group $m\bar{3}m$ outside its subgroup $m\bar{3}$, so
the altermagnetic order parameter transforms as the
one-dimensional representation $A_{2g}$ of $m\bar{3}m$. The
lowest-order axial vector that is odd under time reversal and
linear in this order parameter is the cubic harmonic
\begin{equation}
\mathbf h \propto
\bigl(n_x(n_y^2-n_z^2),\;
      n_y(n_z^2-n_x^2),\;
      n_z(n_x^2-n_y^2)\bigr),
\label{eq:hvec}
\end{equation}
where $\hat{\mathbf n}=(n_x,n_y,n_z)$ is the N\'eel vector.
Equation~\eqref{eq:hvec} vanishes identically for
$\hat{\mathbf n}$ along $\langle100\rangle$ and $\langle111\rangle$
and gives $\mathbf h\propto(0,1,-1)$ for
$\hat{\mathbf n}\parallel[011]$. The same conclusion follows from
the magnetic point groups of the three configurations,
$4'/mm'm$ for $[001]$, $\bar{3}m$ for $[111]$ and $m'm'm$ for
$[011]$: only the last admits a time-reversal-odd axial vector,
directed along its unprimed two-fold axis $[01\bar{1}]$. Its
nonzero projection onto the propagation direction $[001]$
therefore permits helicity-dependent optical transitions in the
geometry considered here, with a magnitude that varies as
$\sin^{2}\theta\cos\theta$ when the N\'eel vector is rotated in
the $(100)$ plane by an angle $\theta$ from $[001]$ toward $[010]$.
Symmetry determines whether such a response is
allowed but not its magnitude; analogous
N\'eel-vector-dependent dichroism has been derived and observed
for MnTe~\cite{Hariki2024xray,Kunes2025neel} and for the insulating
altermagnet hematite~\cite{Hariki2024determination,Pan2026experimental}.

To test these selection rules we performed spin--orbit-coupled noncollinear calculations
for four N\'eel-vector orientations:
$[001]$, $[100]$, $[111]$, and $[011]$.
All calculations employed the same
ambient-pressure structure and computational
parameters, differing only in the orientation
of the compensated magnetic moments.
The transition strengths were evaluated on a $12\times12$ grid in the $k_z=0$ plane.

\begin{figure}[tbp]
\centering
\includegraphics[width=0.7\textwidth]{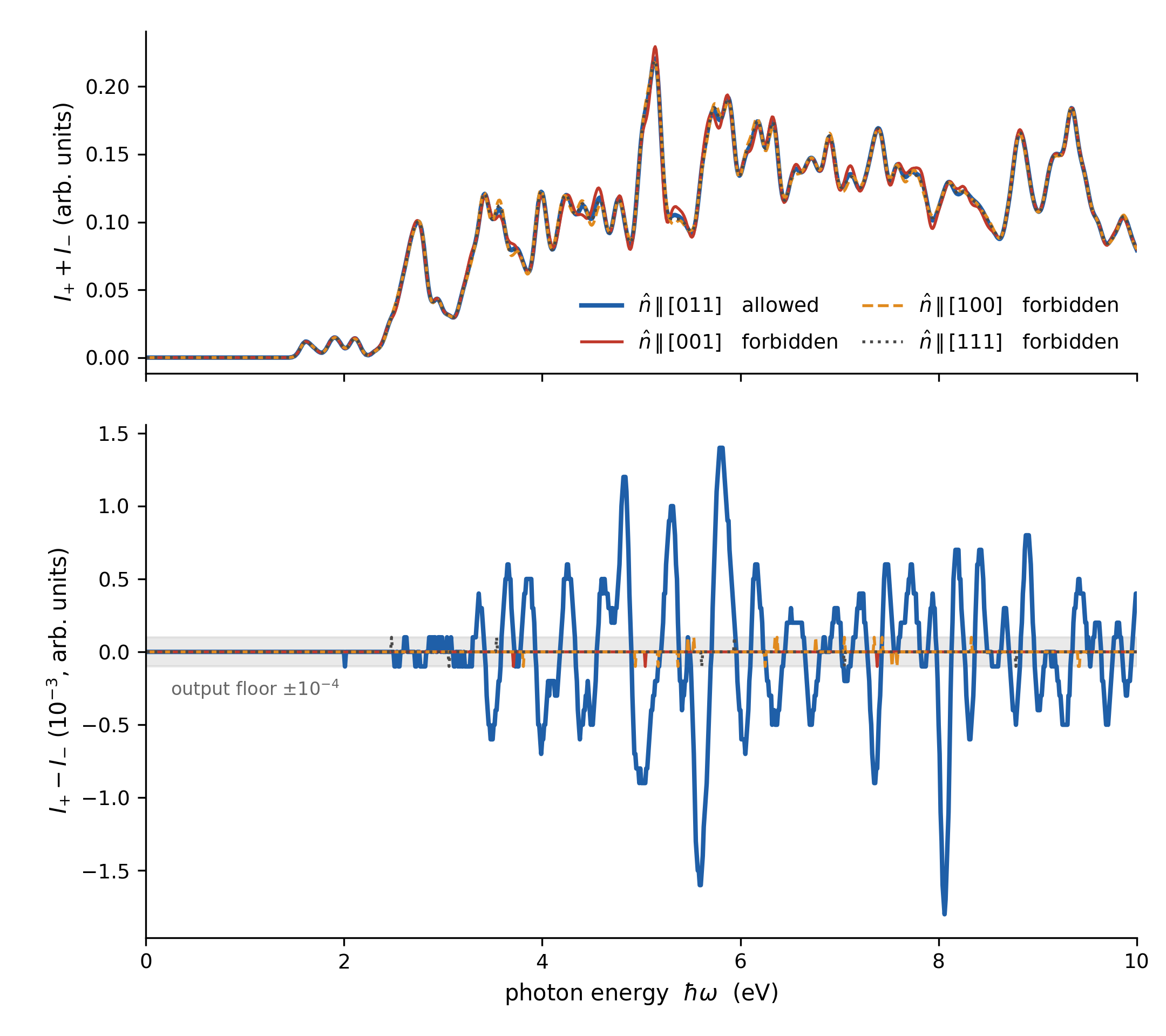}
\caption{\label{fig:mcd}
\textbf{N\'eel-vector dependence of helicity-resolved interband transitions.}
Helicity-resolved interband optical transitions in the
compensated antiferromagnetic configuration
of \gdbc{}, calculated including spin--orbit
coupling for light propagating along [001].
Upper panel: sum of the transition strengths
for opposite circular polarizations,
$I_{+}+I_{-}$.
Lower panel: helicity-dependent difference,
$I_{+}-I_{-}$, for N\'eel vectors along
[001], [100], [111], and [011].
A finite asymmetry is obtained for [011],
whereas the other three orientations
remain at the numerical output floor.
Spectra from a $12\times12$ grid in the $k_z=0$ plane; relative transition strengths, not absorption spectra.
}
\end{figure}

The summed transition strengths are nearly identical for the four orientations (Fig.~\ref{fig:mcd}). The helicity differences are not: a finite asymmetry appears for $\hat{\mathbf n}\parallel[011]$, whereas $[001]$, $[100]$ and $[111]$ stay at the numerical output floor, as the selection rules require.

For $[011]$ the peak contrast is $1.45\%$ of the summed transition strength and the average over the sampled range is $0.28\%$. For the three forbidden orientations the differences stay below the output resolution of about $10^{-4}$.

The magnitude of the contrast depends on the momentum sampling and on the matrix elements, but the on--off pattern across orientations is fixed by symmetry and the calculation reproduces it. Reorienting the N\'eel vector therefore switches helicity-dependent optical transitions without net magnetization.

\subsection*{Thermodynamic and dynamical stability}

We next assess the thermodynamic stability of \gdbc{} under pressure and its dynamical
stability at ambient conditions. The $Pm\bar{3}n$ clathrate structure has been made in
SrB$_3$C$_3$ at high pressure and recovered to ambient
conditions~\cite{Zhu2020SciAdv,Strobel2021lanthanum,Strobel2025typeI}. The same route requires
that the Gd clathrate be stable against every competing phase at the synthesis pressure and
locally stable once the pressure is released.

Pressure favors the clathrate. Its closest competitor at the \gdbc{} composition is a layered
$Amm2$ structure, which is lower in enthalpy at low pressure, by 55~meV/atom at 30~GPa; the
clathrate overtakes it near 62~GPa and is 61~meV/atom lower at 100~GPa in spin-polarized
$r^{2}$SCAN$+U$. The crossover moves only to about 70~GPa without the Hubbard term, which
largely cancels between phases that carry the same half-filled $4f^{7}$ shell. Spin
polarization, by contrast, is essential: at 50~GPa a nonmagnetic treatment favors the
clathrate by 26.8~meV/atom, whereas the spin-polarized one places it 19.8~meV/atom above the
layered phase, because the two structures gain different magnetic stabilization energies.

\begin{figure}[tbp]
\centering
\begin{minipage}[t]{0.50\textwidth}
\centering
\textbf{a}\\[2pt]
\includegraphics[width=\linewidth]{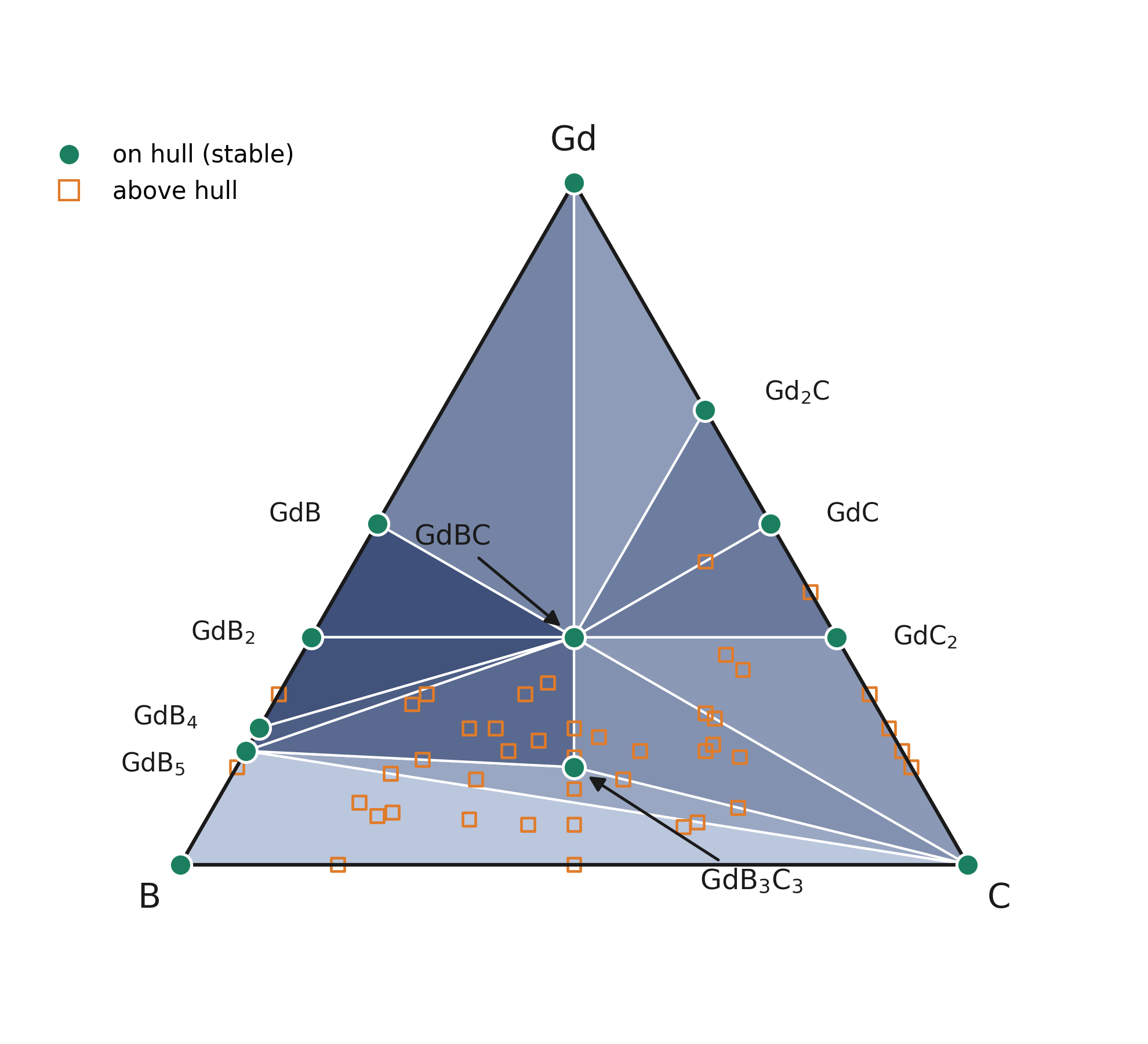}
\end{minipage}\hfill
\begin{minipage}[t]{0.46\textwidth}
\centering
\textbf{b}\\[2pt]
\includegraphics[width=\linewidth]{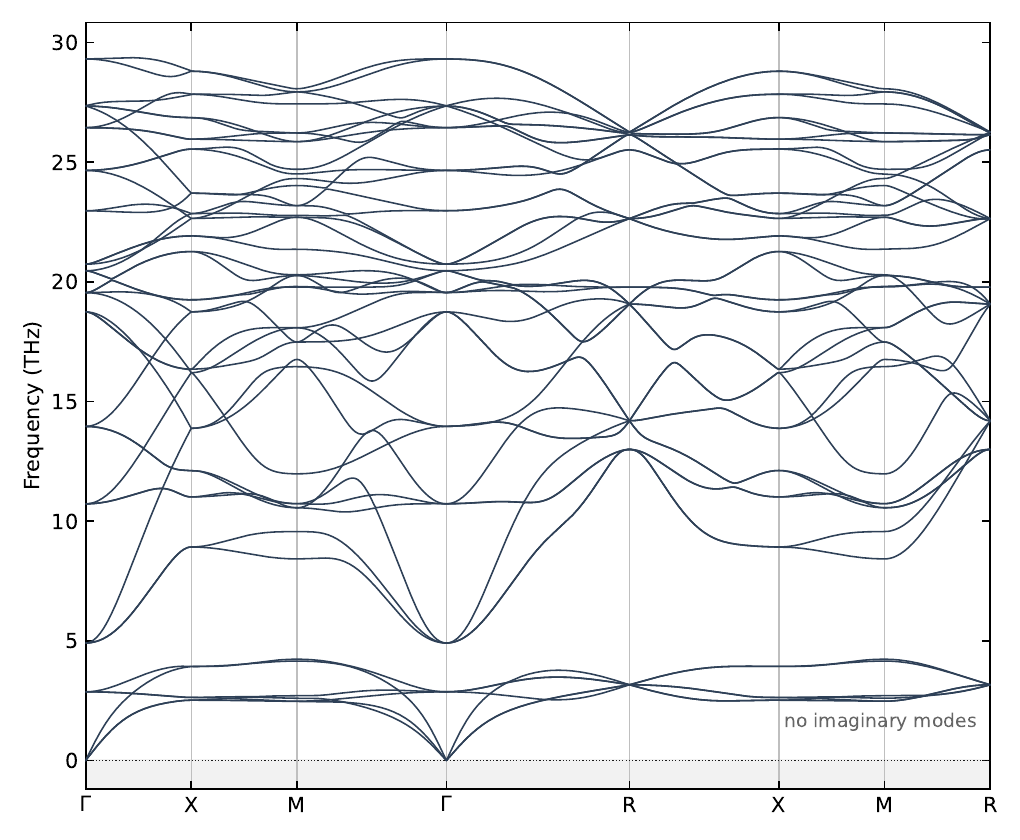}
\end{minipage}
\caption{\label{fig:stability}
\textbf{Thermodynamic and dynamical stability of \gdbc{}.}
\textbf{a}~Calculated Gd--B--C ternary convex hull
at 80~GPa using spin-polarized
$r^{2}$SCAN calculations.
Filled circles denote compositions on
the convex hull, while open squares
represent structures above the hull.
White lines indicate the tie-lines
between stable phases.
The clathrate \gdbc{} lies on the
calculated hull, with GdB$_5$,
GdBC, and elemental carbon defining
its neighboring decomposition facet.
\textbf{b}~Phonon dispersion of the relaxed
antiferromagnetic \gdbc{} structure
at ambient pressure, calculated using
the finite-displacement method in a
$2\times2\times2$ supercell.
All 42 branches have real frequencies along the paths shown.
}
\end{figure}

At 80~GPa the clathrate is stable against decomposition into any other Gd--B--C phase. An
extensive structure search across the ternary system, 228 relaxed structures at 52
compositions with spin-polarized $r^{2}$SCAN enthalpies and the elemental references
monoclinic $C2/m$ Gd, $\gamma$-B$_{28}$ and diamond~\cite{Samudrala2014,Oganov2009}, places
\gdbc{} on the convex hull (Fig.~\ref{fig:stability}a and Supplementary Fig.~3), with GdB$_5$,
GdBC and carbon as its neighboring vertices. The $Pm\bar{3}n$ clathrate is also the
lowest-enthalpy structure found at its own composition, 14.4~meV/atom below the layered phase.

At ambient pressure the clathrate is dynamically stable. The phonon dispersion of the relaxed
antiferromagnetic structure (Fig.~\ref{fig:stability}b) has real frequencies in all 42
branches along the high-symmetry paths. \gdbc{} is therefore thermodynamically stable at the
pressure where this family is made and dynamically stable where it is recovered, which
indicates that it can be synthesized under compression and retained as a metastable phase on
decompression, as SrB$_3$C$_3$ was.

\section*{Discussion}

The boron--carbon clathrates provide a platform in which the
altermagnetic class is set by the host and not by the magnetic ion.
The $2a$ guest site of $Pm\bar{3}n$ lies on an index-two subgroup of
the cubic point group, so the cage exchanges the two guest
sublattices only by rotations. Any compensated guest order is
therefore a cubic $i$-wave altermagnet with six symmetry-enforced
nodal planes.
\gdbc{} is the first example: spin-group analysis and
three-dimensional first-principles calculations give a sixth-order
momentum dependence of the nonrelativistic spin splitting, with
large splitting in the dispersive conduction bands.

The energy scale of the magnetic order is small. The
compensated AFM configuration lies below the FM configuration by
about 1~meV per Gd at $\Ueff = 7$~eV. A mean-field estimate from an
energy difference of this size places the ordering temperature at
roughly 10~K~\cite{Mitra2008magnetic},
comparable to the 15--16~K N\'eel temperature of
GdB$_6$~\cite{Kunii1985gdb6,Amara2005exchange} and well below the
$\sim$70~K of the semiconducting ferromagnet
GdN~\cite{Granville2006semiconducting}; a quantitative transition
temperature would require exchange constants from several ordered
configurations and is not attempted here. Because the two
configurations are so close in energy, modest strain or magnetic
field can be expected to switch between them. The $i$-wave
classification, the nodal planes, the 0.19~eV conduction-band
splitting and the helicity selection rule are properties of the
compensated configuration, whereas the FM state is a conventional
ferromagnetic semiconductor without nodal structure. The weak
anisotropy of the Gd$^{3+}$ moments, in turn, makes the N\'eel
vector of the compensated state easy to rotate.

Experimentally, the most direct signature is spin- and
angle-resolved photoemission of the conduction band, where the
splitting at generic momenta and its vanishing on the $\{110\}$
planes would be observable as it has been in MnTe and
CrSb~\cite{Krempasky2024,Lee2024,Reimers2024,yang2025three}.
The second is optical or x-ray magnetic circular dichroism for light
along $[001]$, which Eq.~\eqref{eq:hvec} predicts to appear only
for N\'eel vectors away from the cubic axes and body diagonals and
to follow a $\sin^{2}\theta\cos\theta$ law, as
demonstrated for MnTe~\cite{Hariki2024xray,Amin2024nanoscale} and
hematite~\cite{Pan2026experimental}. The third is a finite
dichroism together with a vanishing net moment in magnetometry,
which distinguishes the compensated from the ferromagnetic
state~\cite{Jungwirth2026signatures}.

The argument extends beyond \gdbc{}. The symmetry argument applies
unchanged to every rare-earth or transition-metal guest, so the
lanthanide series $M$B$_3$C$_3$ (Supplementary Table~1) is a family
of candidates that differ only in their magnetic ground state and
exchange energy. A divalent $4f^{7}$ guest such as Eu$^{2+}$, the ion
of the ferromagnetic clathrate
Eu$_8$Ga$_{16}$Ge$_{30}$~\cite{Paschen2001eu8ga16ge30}, offers a
different exchange balance, and the type-I clathrate adds a second,
large-cage guest site~\cite{Strobel2025typeI}. Because the wave class
is fixed by the framework symmetry, a few per cent of tetragonal or
rhombohedral strain should convert the $i$-wave splitting into a
$d$- or $g$-wave one (Supplementary Note~4). The lower-order form
factor is far larger near the zone center, so a single material would
host a strain-switchable altermagnetic order. Finally, the weak
anisotropy of a half-filled $4f$ shell makes the N\'eel vector
rotatable and, through the magnetic point group, switches the
helicity-resolved optical response on and off. Ordered nonmagnetic frameworks with $L=0$ guests therefore give symmetry-controlled optical response without net magnetization. The
predicted thermodynamic stability under pressure and dynamical
stability at ambient conditions, in a family already synthesized by
this route, motivate the experimental test.

\FloatBarrier
\section*{Methods}

\subsection*{First-principles calculations}
First-principles calculations were performed using the
projector-augmented-wave method~\cite{Blochl1994}
as implemented in VASP 6.4.2~\cite{Kresse1996}. The exchange-correlation
energy was described by the $r^{2}$SCAN meta-GGA
functional~\cite{Furness2020}, with a plane-wave energy cutoff of
520~eV and an electronic convergence threshold of $10^{-6}$~eV.
The Gd $4f$ electrons were explicitly
included in the valence states, and their on-site Coulomb
interactions were treated using the DFT$+U$
approach~\cite{Anisimov1991band} in the Dudarev
form~\cite{Dudarev1998} with an effective Hubbard parameter
$\Ueff = 7.0$~eV. Total-energy calculations
employed a $9\times9\times9$ Monkhorst--Pack $k$-point
mesh~\cite{Monkhorst1976special}, with ferromagnetic and
antiferromagnetic configurations initialized independently under
otherwise identical computational conditions.

Cold-started $r^{2}$SCAN$+U$ calculations of partially filled $4f$ shells can diverge
or settle into unphysical $4f$ occupations. A chained PBE$+U \rightarrow r^{2}$SCAN$+U$
restart, or a ramp in $U$, was therefore used throughout the lanthanide screen and for
the antiferromagnetic configurations that did not converge from a cold start. All calculations used \texttt{ALGO}~$=$~All,
\texttt{LASPH}~$=$~\texttt{.TRUE.}, Gaussian smearing with $\sigma = 0.05$~eV, and
mixing parameters \texttt{AMIX}/\texttt{BMIX}~$= 0.1/0.5$ with
\texttt{AMIX\_MAG}/\texttt{BMIX\_MAG}~$= 0.1/0.5$. \texttt{LMAXMIX}~$=6$ was set
throughout, as required for $f$-electron charge-density mixing. The Hubbard correction
used \texttt{LDAUL}~$=3$ and, for the production runs, \texttt{LDAUTYPE}~$=2$ with
\texttt{LDAUJ}~$=0$.

\subsection*{Hubbard $U$ from constrained RPA}
The Hubbard parameters of the Gd $4f$ shell were computed with the
constrained random-phase
approximation~\cite{Aryasetiawan2006calculations}
as implemented in VASP, screening the bare Coulomb interaction by all
polarization channels except those within the $4f$ target manifold
(14 $f$ projectors, seven per Gd site), on a spin-polarized PBE$+U$
reference state (cutoff 520~eV, 256 bands, $4\times4\times4$ mesh,
response cutoff 200~eV, static limit). Because the screened $U$
depends on the $U$ used to build the reference, the calculation was
iterated to self-consistency:
$U_{\mathrm{out}} = 4.09$, 6.60, 7.44 and 7.51~eV for
$U_{\mathrm{in}} = 0$, 4, 7 and 7.5~eV, with $J = 0.23$~eV
throughout and a bare on-site $U$ of 28~eV, giving the fixed point
$U^{\ast} = 7.51$~eV, $J^{\ast} = 0.232$~eV and
$\Ueff^{\ast} = U^{\ast}-J^{\ast} = 7.28$~eV. A nonmagnetic reference
yields a negative screened $U$ because the half-filled $4f$
manifold is then pinned at the Fermi level, which is why the
spin-polarized reference is required.

\subsection*{Hexaboride validation}
As an empirical check of the functional and $\Ueff$, FM and A-type
AFM configurations of GdB$_6$ (space group $Pm\bar{3}m$,
$a = 4.108$~\AA, cell doubled along $c$, $8\times8\times4$ mesh,
cutoff 520~eV, Methfessel--Paxton smearing of 0.1~eV, otherwise
identical settings) were compared at $\Ueff = 5$--8~eV. The
two-sublattice A-type arrangement serves as a proxy for the more
complex experimental ordering of
GdB$_6$~\cite{Luca2004neutron,Amara2005exchange}. The AFM state is
lower by 13--14~meV per Gd for $\Ueff = 5$--7~eV (Supplementary
Table~3). The ferromagnet EuB$_6$~\cite{Sullow1998structure}, treated
in the same way, is placed within 1~meV per Eu of degeneracy with
the opposite sign, which we take as the intrinsic resolution of
this level of theory for carrier-mediated exchange.

\subsection*{Band structure and spin splitting}
Electronic band structures and momentum-dependent spin splitting
$\Delta(\kv)$ were evaluated using zero-weight $k$ points appended
to the self-consistent mesh. These included the high-symmetry
path $\Gamma$--X--M--$\Gamma$--R--X$|$M--R with 60 points per segment,
a $31\times31$ grid in the $k_z=0$ plane, and a
$15\times15\times15$ grid covering the three-dimensional
Brillouin zone. The three-dimensional scan was run in five blocks, each pairing 125
weighted self-consistent points with 675 zero-weight grid points, because the
monolithic calculation exceeded available memory.
Crystal symmetry and magnetic-sublattice
connectivity were analyzed using spglib~\cite{Togo2024spglib}. The
framework-generality and strain statements of Supplementary Note~4
were obtained by enumerating, for each (strained) structure with a
two-sublattice N\'eel configuration on the guest sites, the
space-group operations that exchange the two sublattices and
classifying the lowest-degree even polynomial in $\mathbf{k}$ that is
odd under every such operation and even under every
sublattice-preserving one; degree 2, 4 and 6 correspond to $d$-, $g$-
and $i$-wave splitting.

\subsection*{Noncollinear calculations and optical transitions}
Noncollinear calculations incorporating
spin--orbit coupling were performed with
\texttt{vasp\_ncl}~\cite{Hobbs2000fully},
restarted from the converged collinear antiferromagnetic density,
to investigate magnetocrystalline anisotropy and helicity-dependent optical
transitions, with the N\'eel-vector orientation specified
through \texttt{SAXIS}. The anisotropy and the SOC band comparison
of Supplementary Fig.~2 used the full zone with the N\'eel vector
along [001], [100] and [110], at the lattice constant
$a = 4.8495$~\AA{} used in an earlier version of the lanthanide screen; the
$\mu$eV-scale anisotropy is not expected to change at 4.8175~\AA. The
magneto-optical calculations used $\mathrm{NBANDS} = 160$ and a
$12\times12$ mesh over the $k_z = 0$ plane, with the four runs differing only in
\texttt{SAXIS}. The transition matrix elements were evaluated
using the pseudo-wavefunction momentum operator in the plane-wave
representation, without the projector-augmented-wave augmentation
contributions~\cite{Gajdos2006linear}.

\subsection*{Phonons}
Phonon dispersions were calculated using the finite-displacement
method in a $2\times2\times2$ supercell with
phonopy~\cite{Togo2015}.

\subsection*{Structure search and convex hull}
Thermodynamic stability was investigated by constructing the
Gd--B--C convex hull at 80~GPa. Candidate structures were
generated using CALYPSO~\cite{Wang2012} at the PBE level,
followed by spin-polarized $r^{2}$SCAN relaxation of the three
lowest distinct structures from each search, yielding
228 structures across 52 compositions.
Re-relaxations used \texttt{ENCUT}~$= 600$~eV, \texttt{ISIF}~$=3$,
\texttt{PSTRESS}~$= 800$~kbar, \texttt{NSW}~$= 400$, \texttt{ISPIN}~$=2$ and a
$k$-spacing of 0.030.
The electronic, magnetic, optical and phonon properties were evaluated at ambient pressure at the relaxed lattice constant $a=4.8175$~\AA, which corresponds to a recovered clathrate phase, following the high-pressure synthesis and ambient recovery of SrB$_3$C$_3$~\cite{Zhu2020SciAdv}.

\subsection*{Use of AI tools}
Large language models (Claude, Anthropic; ChatGPT, OpenAI) were used under the direction of
the authors to proofread text, to write and check analysis and plotting scripts, and to audit
the manuscript for internal consistency. All calculations, results, interpretations and
conclusions are those of the authors; every AI-assisted passage and script was verified by the
authors, who take full responsibility for the content.

\section*{Data availability}
The data that support the findings of this study are available from the corresponding
author upon reasonable request.

\section*{Code availability}
VASP is a commercial code; CALYPSO, phonopy and spglib are available from their
developers. Analysis scripts are available from the corresponding author on request.

\section*{Acknowledgements}
This work was supported by the faculty funds of the Office of the Dean of the School of Arts
and Sciences--Newark (SASN) of Rutgers University--Newark. Computational resources were
provided by the Amarel cluster of the Rutgers Office of Advanced Research Computing (OARC).

\section*{Author contributions}
L.Z. conceived, designed and directed the research. M.W. performed the calculations and
drafted the manuscript. All authors edited the manuscript.

\section*{Competing interests}
The authors declare no competing interests.

\bibliographystyle{unsrtnat}
\bibliography{refs,refs_additions}

\clearpage
\setcounter{figure}{0}\setcounter{table}{0}\setcounter{equation}{0}
\renewcommand{\figurename}{Supplementary Figure}\renewcommand{\tablename}{Supplementary Table}
\renewcommand{\thefigure}{\arabic{figure}}\renewcommand{\thetable}{\arabic{table}}

\begin{center}{\Large\textbf{Supplementary Information}}\end{center}

\vspace{1em}

\section*{Supplementary Note 1: Crystal structure}

\begin{figure}[h]
\centering
\includegraphics[width=0.5\textwidth]{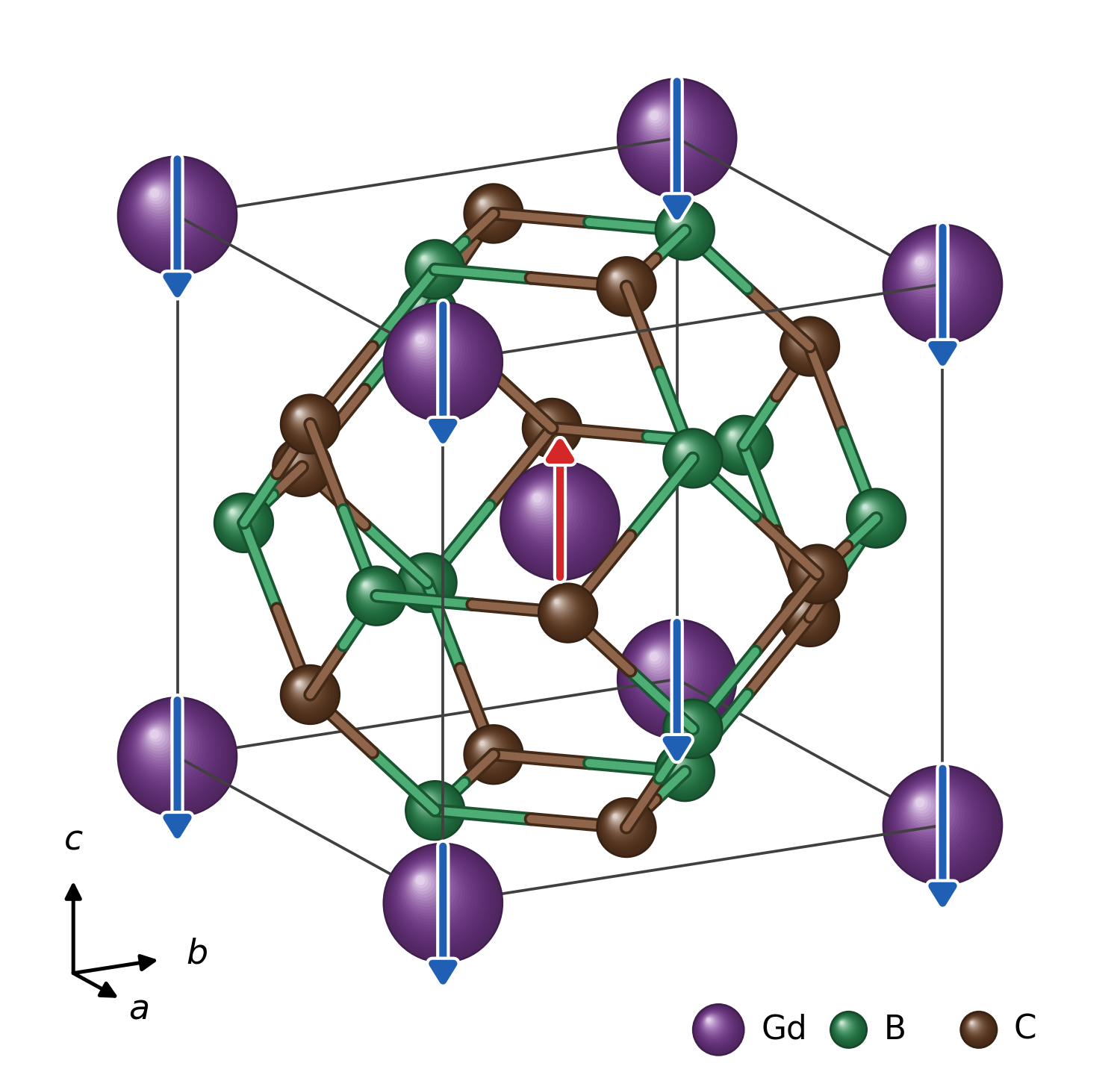}
\caption{\label{si:fig:structure}
\textbf{Crystal structure of \gdbc{}.}
Conventional cell of $Pm\bar{3}n$ \gdbc{} ($a=4.8175$~\AA). Gd (purple) at $2a$, $(0,0,0)$
and $(\tfrac12,\tfrac12,\tfrac12)$; B (green) at $6c$ and C (brown) at $6d$. The B--C bonds
(1.70~\AA) of the B$_{12}$C$_{12}$ cage around the body-center Gd are drawn. Arrows mark the
antiparallel moments of the two Gd sublattices in the compensated ground state.}
\end{figure}

Supplementary Fig.~\ref{si:fig:structure} shows the conventional cell of \gdbc{} with the
compensated N\'eel order used throughout.

\section*{Supplementary Note 2: Lanthanide screen}

Supplementary Table~\ref{si:tab:screen} lists the ferromagnetic--antiferromagnetic energy
difference of the lanthanide members $M$B$_3$C$_3$ at three values of the effective Hubbard
parameter, $\Ueff = 4.2$, 5.5 and 7.0~eV, under otherwise identical conditions. No single,
element-specific $U$ is assumed: the purpose of the table is to establish whether the
ground state of each member is stable across a range of $\Ueff$ that brackets the values
appropriate to trivalent $4f$ states in an insulating host. Because the altermagnetic
eligibility and the $i$-wave class are fixed by the framework (Supplementary Note~4), any
member with a compensated ground state would be a cubic $i$-wave altermagnet.

The pattern is uniform. For every member that converges reliably, increasing $\Ueff$ moves
$\Delta E$ towards the compensated state, but Gd is the only member that crosses: it is
ferromagnetic at 4.2 and 5.5~eV and compensated at 7.0~eV, with the crossover at
$\Ueff \approx 6.1$~eV (Supplementary Table~\ref{si:tab:uscan}), below its first-principles value
$\Ueff^{\ast} = 7.28$~eV (Methods). All other reliable entries are ferromagnetic at every
$\Ueff$ computed. Ce, Pr and Er, however, remain within 0.9~meV of degeneracy throughout
and should be regarded as nearly degenerate rather than ferromagnetic. Sm and Eu
are the only members whose $\Delta E$ is not monotonic in $\Ueff$; their $4f$ occupation lies
between the trivalent and divalent values (5.48 and 6.88 electrons at 4.2~eV) and shifts
with $\Ueff$, so that the valence itself changes across the range. All members are
evaluated at the lattice constant of \gdbc{}, so the table is a ranking under common
conditions rather than a determination of each ground state.

\begin{table}[h]
\centering
\caption{\label{si:tab:screen}
\textbf{Magnetic ordering energy of the lanthanide clathrates $M$B$_3$C$_3$ as a function of
the Hubbard parameter.}
$\Delta E = E_{\mathrm{FM}} - E_{\mathrm{AFM}}$ (meV per magnetic unit cell, two $M$; same
sign convention as Supplementary Table~\ref{si:tab:uscan}, positive $\Delta E$ meaning the compensated
antiferromagnetic state is lower). All entries: collinear $r^{2}$SCAN$+U$ without
spin--orbit coupling in the Dudarev form, as in the production calculations, at the lattice
constant of \gdbc{}, $a = 4.8175$~\AA, on a $5\times5\times5$ Monkhorst--Pack mesh, with the
$4f$ electrons in the valence and initial moments equal to the spin-only Hund value of
$M^{3+}$. The Gd entry at 7.0~eV reproduces the production calculation on the same mesh to
0.001~meV. Every entry was checked both for self-consistency and for a physical $4f$
occupation and moment on each site.
$^{\ast}$Self-consistency stopped at the 300-step limit with a last-step energy change at
least an order of magnitude below $|\Delta E|$; the sign is reliable.
$^{\dagger}$Not reliable: the antiferromagnetic calculation did not converge and its energy
changed discontinuously between 5.5 and 7.0~eV, indicating a different $4f$ occupation.
--~For Tb, Dy and Ho at $\Ueff \ge 5.5$~eV the self-consistent cycle left the physical
subspace: the projected $4f$ occupation rose above the 14 electrons a $4f$ shell can hold
(Tb, Dy) or the two guest sites became inequivalent (Ho). This is a known instability of
DFT$+U$ for partially filled $f$ shells at large $U$, which would require control of the
$4f$ occupation matrix; at 4.2~eV all three converge to physical states. Yb converges to
nonmagnetic Yb$^{2+}$ ($4f^{14}$) and La ($4f^{0}$) is nonmagnetic; neither is listed.}
\vspace{4pt}
\begin{tabular}{lcrrrl}
\toprule
$M$ & $4f^{n}$ ($M^{3+}$) & \multicolumn{3}{c}{$\Delta E$ (meV per cell)} & ground state \\
\cmidrule(lr){3-5}
 & & $\Ueff = 4.2$ & 5.5 & 7.0~eV & \\
\midrule
Ce & $4f^{1}$  & $-0.89$          & $-0.62$          & $-0.43$         & FM (near-degenerate) \\
Pr & $4f^{2}$  & $-0.26$          & $-0.09$          & $-0.05$         & FM (near-degenerate) \\
Nd & $4f^{3}$  & $-1.72$          & $-1.25$          & $-0.92$         & FM \\
Pm & $4f^{4}$  & $-4.38$          & $-2.94$          & $-1.42$         & FM \\
Sm & $4f^{5}$  & $-12.7^{\ast}$   & $-14.2^{\ast}$   & $-10.8^{\ast}$  & FM \\
Eu & $4f^{6}$  & $-6.95$          & $-11.76$         & $-6.99$         & FM \\
Gd & $4f^{7}$  & $-2.71$          & $-1.05$          & $+2.01$         & FM $\rightarrow$ AFM at $\Ueff \approx 6.1$~eV \\
Tb & $4f^{8}$  & $-0.17^{\ast}$   & --               & --              & FM (near-degenerate) at 4.2~eV \\
Dy & $4f^{9}$  & $-3.68^{\ast}$   & --               & --              & FM at 4.2~eV \\
Ho & $4f^{10}$ & $-21.5^{\ast}$   & --               & --              & FM at 4.2~eV \\
Er & $4f^{11}$ & $-0.30$          & $-0.20$          & $-0.19$         & FM (near-degenerate) \\
Tm & $4f^{12}$ & $-4.11$          & $-2.31$          & $(+11.9)^{\dagger}$ & FM at 4.2--5.5~eV \\
\bottomrule
\end{tabular}
\end{table}

\FloatBarrier
\section*{Supplementary Note 3: Dependence of the magnetic ordering energy on $U$}

Supplementary Table~\ref{si:tab:uscan} lists the ferromagnetic--antiferromagnetic energy
difference of \gdbc{} as a function of the effective Hubbard parameter in the
$r^{2}$SCAN$+U$ (Dudarev) calculations described in the Methods. The crossover from
ferromagnetic to compensated antiferromagnetic order lies at $\Ueff \approx 6.1$~eV.
Supplementary Table~\ref{si:tab:hexa} gives the hexaboride validation of the functional
and $\Ueff$.

\begin{table}[h]
\centering
\caption{\label{si:tab:uscan}
\textbf{Magnetic ordering energy versus $\Ueff$.}
$\Delta E = E_{\mathrm{FM}} - E_{\mathrm{AFM}}$ for \gdbc{} as a function of the
effective Hubbard parameter ($r^{2}$SCAN$+U$, Dudarev scheme, $9\times9\times9$ mesh,
$a = 4.8175$~\AA, cutoff 520~eV). Positive $\Delta E$
means the compensated antiferromagnetic state is the ground state. Energies are per
magnetic unit cell (two Gd).}
\vspace{4pt}
\begin{tabular}{ccl}
\toprule
$\Ueff$ (eV) & $\Delta E$ (meV per cell) & ground state \\
\midrule
5.0  & $-1.74$  & FM \\
6.0  & $-0.17$  & FM (nearly degenerate) \\
7.0  & $+1.99$  & AFM \\
8.0  & $+4.90$  & AFM \\
9.0  & $+8.81$  & AFM \\
10.0 & $+14.13$ & AFM \\
\bottomrule
\end{tabular}
\end{table}

\begin{table}[h]
\centering
\caption{\label{si:tab:hexa}
\textbf{Hexaboride validation.}
$E_{\mathrm{AFM}} - E_{\mathrm{FM}}$ (meV per rare-earth ion; negative = A-type AFM
lower) for GdB$_6$ and EuB$_6$ in the $1\times1\times2$ cell described in the Methods
($r^{2}$SCAN$+U$, Dudarev). Experimentally GdB$_6$ is an antiferromagnet
($T_{\mathrm N} \approx 15$--16~K) and EuB$_6$ a ferromagnet
($T_{\mathrm C} \approx 15$~K). The GdB$_6$ run at $\Ueff = 8$~eV converged into the
FM basin from the AFM start and is omitted.}
\vspace{4pt}
\begin{tabular}{lcccc}
\toprule
compound & $\Ueff = 5$ & 6 & 7 & 8~eV \\
\midrule
GdB$_6$ & $-13.12$ & $-13.48$ & $-13.68$ & -- \\
EuB$_6$ & $-0.76$ & $-0.53$ & $-0.44$ & $-0.40$ \\
\bottomrule
\end{tabular}
\end{table}

\FloatBarrier
\section*{Supplementary Note 4: Framework symmetry and strain retuning of the
altermagnetic class}

In $Pm\bar{3}n$ the guest occupies Wyckoff site $2a$, whose site-symmetry group is
$m\bar{3}$ ($T_h$, order 24), an index-two subgroup of the point group $m\bar{3}m$
($O_h$, order 48). The 24 space-group operations that exchange the two guests are the
coset $O_h\setminus T_h = \{6C_4,\,6C_2',\,6S_4,\,6\sigma_d\}$, each carrying the
fractional translation $(\tfrac12,\tfrac12,\tfrac12)$; the identity and the inversion
lie in $T_h$ and preserve each sublattice. The eligibility of the two-sublattice
N\'eel state for altermagnetism is therefore a consequence of the site symmetry and
holds for every guest element, lattice parameter and cage decoration. The
lowest-degree even polynomial in $\mathbf{k}$ that is odd under every
sublattice-exchanging operation and even under every sublattice-preserving one is
of degree six, $F_6(\mathbf{k}) = (k_x^2-k_y^2)(k_y^2-k_z^2)(k_z^2-k_x^2)$, which
identifies the $i$-wave class; under a tetragonal or rhombohedral distortion the
sublattice-exchanging set shrinks to a subgroup and the lowest admissible degree
drops to two or four (Supplementary Table~\ref{si:tab:strain}). Known altermagnets
treated with the same procedure return their accepted classes (RuO$_2$, $d$-wave;
MnTe, $g$-wave), and the A-type order of a hexaboride lattice in a doubled cell is
correctly identified as a conventional, translation-connected antiferromagnet.

\begin{table}[h]
\centering
\caption{\label{si:tab:strain}
\textbf{Altermagnetic class of the two-sublattice N\'eel state on the clathrate guest
site under symmetry-lowering strain.} Space groups from spglib; ``degree'' is the lowest
degree of the spin-splitting form factor allowed by the sublattice-exchanging
operations. Strains of 2\% are used to define the symmetry; the classification does
not depend on the magnitude.}
\vspace{4pt}
\begin{tabular}{lllcl}
\toprule
structure & distortion & space group & degree & class \\
\midrule
$M$B$_3$C$_3$, type-VII & none (cubic) & $Pm\bar{3}n$ (223) & 6 & $i$-wave \\
 & tetragonal, $\varepsilon_{c} = -2\%$ & $P4_2/mmc$ (131) & 2 & $d$-wave \\
 & biaxial in-plane, $\varepsilon_{ab} = +2\%$ & $P4_2/mmc$ (131) & 2 & $d$-wave \\
 & rhombohedral along $[111]$, 2\% & $R\bar{3}c$ (167) & 4 & $g$-wave \\
type-I B--C clathrate, $2a$ guests & none (cubic) & $Pm\bar{3}n$ (223) & 6 & $i$-wave \\
\bottomrule
\end{tabular}
\end{table}

\FloatBarrier
\section*{Supplementary Note 5: Effect of spin--orbit coupling on the band structure}

\begin{figure}[h]
\centering
\includegraphics[width=0.75\textwidth]{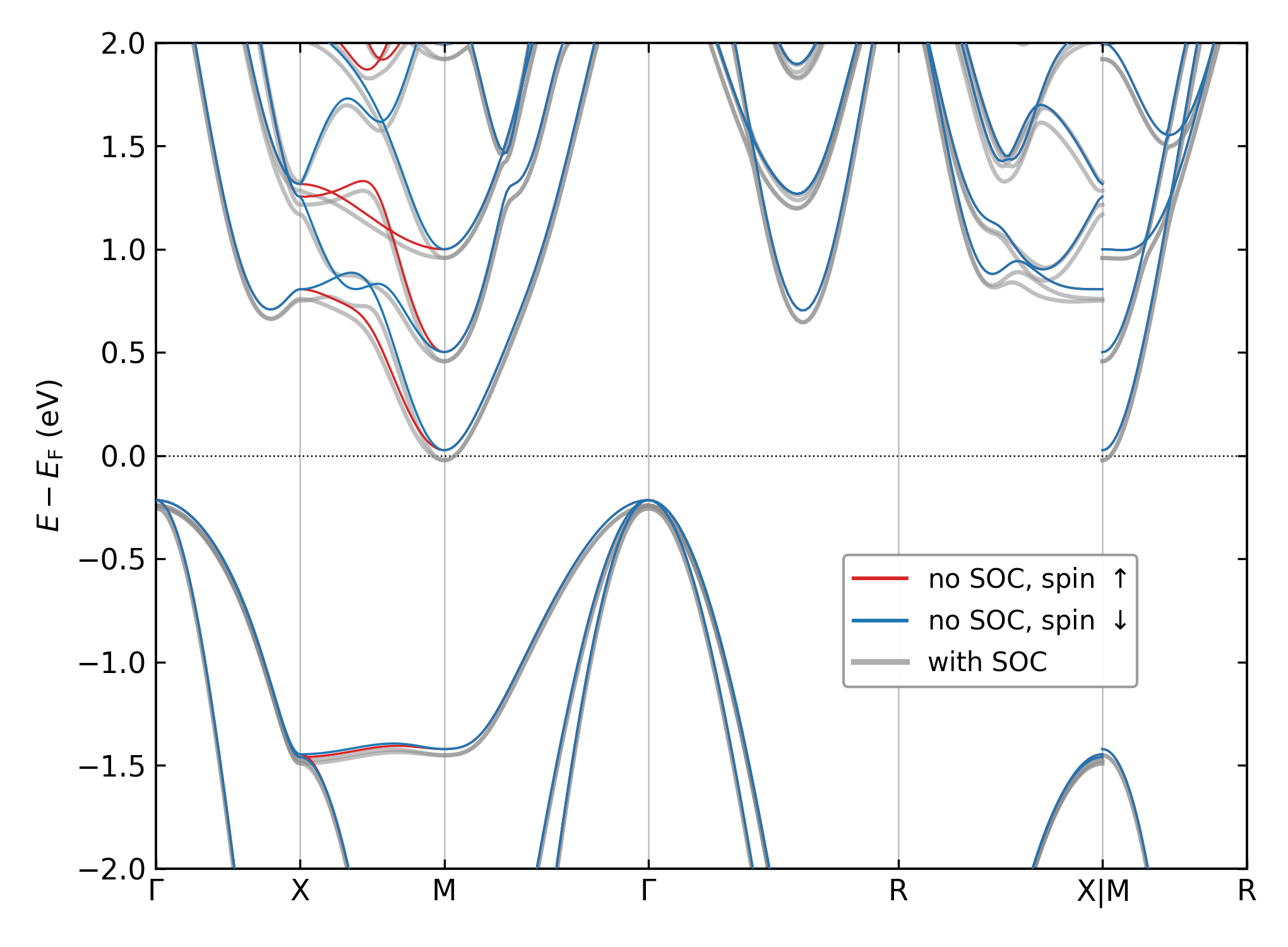}
\caption{\label{si:fig:soc}
\textbf{Band structure with and without spin--orbit coupling.}
Electronic band structures of the compensated antiferromagnetic configuration calculated
with and without spin--orbit coupling, shown within $\Ef\pm2$~eV. The overall band
dispersions near the insulating gap remain nearly unchanged upon inclusion of SOC.}
\end{figure}

Supplementary Fig.~\ref{si:fig:soc} compares the band structure of the compensated
antiferromagnetic configuration calculated with and without spin--orbit coupling
(at $a = 4.8495$~\AA, see Methods). The
dispersions near the insulating gap are nearly unchanged, consistent with the
$L=0$ character of the Gd$^{3+}$ $4f^{7}$ shell discussed in the main text.

\FloatBarrier
\section*{Supplementary Note 6: Full Gd--B--C convex hull at 80~GPa}

\begin{figure}[h]
\centering
\includegraphics[width=0.75\textwidth]{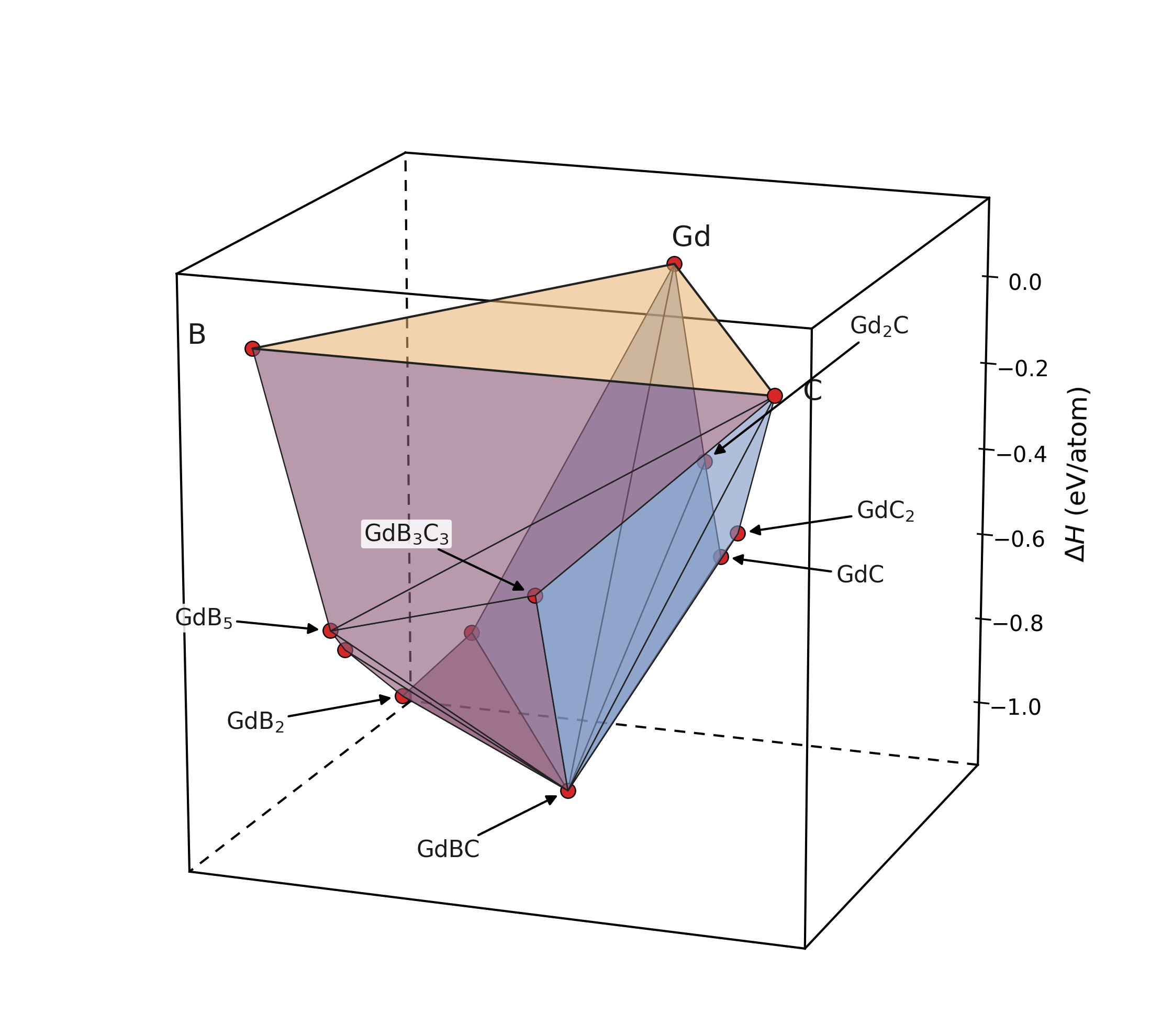}
\caption{\label{si:fig:hull3d}
\textbf{Lower convex hull of the Gd--B--C system at 80~GPa.}
Formation enthalpy is the vertical axis and the $\Delta H = 0$ element triangle is the
upper lid. Red points are the twelve hull vertices -- the three elements together with
Gd$_2$C, GdC, GdC$_2$, GdB, GdB$_2$, GdB$_4$, GdB$_5$, GdBC and \gdbc{}. The 228 refined
structures that lie above the hull are omitted for clarity; their distances above it
are discussed in the main text.}
\end{figure}

Supplementary Fig.~\ref{si:fig:hull3d} shows the lower convex hull of the Gd--B--C system at
80~GPa from which the ternary diagram of Fig.~5a of the main text is projected.

\FloatBarrier

\end{document}